# Evaluating the Impact of Pair Documentation on Requirements Quality and Team Productivity

Nosheen Qamar[1], Nosheen Sabahat[2], Amir Mashmool[3] and Amir Mosavi[4]

[1]Department of Software Engineering, University of Management and Technology, Lahore, 54000, Pakistan

[2]Department of Computer Science, Forman Christian College University, Lahore, 54000, Pakistan

[3]Department of Computer Science, Bioengineering, Robotics and System Engineering, University of Genova, Genova, 16126, Italy

[4]Obuda University, Budapest,1034, Hungary

**Abstract:** The most important deliverable of the requirements engineering process is the software/system requirements specification (SRS) document. Requirements documentation is important during the complete software development lifecycle to share the vision and effective communication between major stakeholders. The Standish Group reported that the top factors behind project failures are related to requirements. By giving the right level of attention to key requirements good quality software can be produced. Therefore, more research is needed in this area and this study is trying to fill this gap. This empirical study aims to examine the importance of pair documentation. Unconventional documentation refers to the approach when two persons work on the same document's requirements collaboratively just like pair programming on the requirements quality and team productivity. Twenty pairs of documentation writers worked into two groups – one group using pair documentation, i.e., the experimental group, and the other one using conventional documentation i.e., the control group. The resultant requirements' documents for the same project, produced by both groups were then compared. It is observed that there is a significant improvement in the quality and productivity of the experimental group using pair documentation. The findings of this study may assist requirement engineers in forming efficient teams that can create high-quality SRS documents.

## 1 Introduction

Over the past few years, software development has become an integral part of the software industry which is dynamic and highly complex. Software is the most important component of almost every field and the success of every business depends on the usage of successful software[1]. However, even though the software has become ubiquitous, there is no key to success in the development of high-quality software. Factors like high-end resources, a better work environment, technical expertise, the right software development process, growing complexities, market trends, changing technology, etc. are all influencing the success of the software[1]. There are various challenges faced by software development teams. For instance, the success of software nowadays is highly dependent on the methodologies used for software development. Furthermore, there is a dire need of integrating various programs, within a limited budget and timelines to follow. In such scenarios, agile methodologies are considered to be more successful and popular in improving software quality [2].

Requirements Engineering (RE) is the first phase in the traditional software development process whereas,

in agile development, it continues throughout the process [3]. The importance of software requirement specification (SRS) lies in documenting the essential requirements of software and its external interface. While many studies have examined the quality of SRS, they often lack proper organization and representation of functional requirements. In one of their research, Stephen et. al evaluated the quality of SRS based on the properties of completeness, correctness, preciseness, and consistency [4]. The analysis of software requirement specification (SRS) documentation requires the extraction of accurate information, which can be challenging as the complexity of software systems continues to increase. Tasks that target specific types of information within the SRS require experts in the field, and large-scale annotation tasks can be costly and overwhelming for a limited number of experts. Failure to manage data effectively can negatively impact operations within an organization [5].

Recent research has found that development and maintenance teams often neglect proper documentation of software requirement specifications due to tight schedules or resource constraints. This can lead to difficulties in maintaining the software system during the maintenance phase. To address this issue, a mechanism is needed to automatically detect and update the software requirement specification documentation immediately after any update is made to the software system. Bhatia et. Al [6] has proposed an ontology-driven software development approach for the automatic detection and updating of software requirement specifications. Although it is common for SRS to contain ambiguities, eliminating them is crucial and can greatly impact the success of the software development process. Ribeiro et. Al [7] has discussed ways to prevent, detect, and fix ambiguities in requirements specifications. It was found that ambiguities did not cause any major problems during development and were resolved through normal discussions and inspections. In one of their research, Zhi et. al [8] suggested a new way to find ambiguities in SRS. The method is called value-oriented review (VOR) and involves identifying core values based on the SRS and finding defects that go against these values. This method helped reviewers find defects that are harder to detect, such as omissions in the SRS. To avoid such ambiguities and mistakes, it is important to have clear requirements for software development. Merugu et. al [9] have suggested using automated tools to analyze requirements and save time and effort. Another research was conducted by Qamar et al [10]. also shows the importance of the RE process by surveying that many fresh graduates have a deep interest in the field of RE and keep learning more about it. The success of a project is highly dependent on the RE process, the outcome of which is an official requirements document. Literature reveals that an unsuccessful RE process is due to multiple reasons like lack of user involvement, missing or incomplete requirements, vague, ambiguous, changing requirements, etc. This official Software Requirements Specification (SRS) document is a contract among the stakeholders[11].

The success of a project is highly dependent on the outcome of the RE process i.e., good quality SRS documents [11]. The studies conducted by Standish Group [12] found that the project success rate has been 29% or less in the past twenty years. The top factors behind it are related to requirements such as lack of user involvement, incomplete/changing requirements & specification, and lack of clear statement of requirements. A project can only be successful if it is given the right level of attention to clarify the key requirements. It may take from 25% to 50% of the total project time [13, 14] to develop a good quality SRS document. Therefore, more research is needed in the area of requirements engineering to make this process more effective and can produce improved quality SRS documents. Our research is trying to fill this gap by proposing a new approach to the software requirements documentation process. The term "Pair Documentation" was first coined by Scott Ambler in his book on Agile Modelling [15]. In pair documentation, two persons work simultaneously on the same requirements document. This concept is similar to the well-known concept of pair programming. This study empirically assesses the usefulness of pair documentation over conventional documentation on the same project. An experiment is designed and conducted using a control group and an experimental group, each consisting of 10 teams working on the requirements document. The experimental group teams work using a pair documentation approach whereas the control group teams work using conventional documentation. The resultant SRS of the projects is then assessed based on the quality and productivity metrics of the two groups of teams.

The rest of the paper is structured as follows. The next section summarizes related work done in this area. Section 3 describes our proposed approach and a detailed description of the design. Section 4 presents important threats to the validity of our research. Finally, Section 5 summarizes the major contributions of this research and presents some directions for future work.

## 2 Literature Review

In the process of automating and developing software, the first and most important step is to identify the requirements and document them in the correct formulation of software requirement specification (SRS). The effectiveness of the various steps in software development depends on the quality of the requirements specification. Thus, the software's ultimate merit depends on the quality of SRS. Therefore, it is important to assess the quality of SRS [16]. Retrieving and extracting information from the SRS is critical in the development of a software product line. Although Natural Language Processing (NLP) techniques have been suggested as a semi-automatic way of optimizing requirement specifications, they have not been widely adopted [17]. Moreover, assessing the quality of Software Requirement Specification (SRS) automatically is challenging, as it requires advanced algorithms to extract features, interpret context, formulate metrics, and document shortcomings [18]. The accuracy, consistency, and completeness of software quality are dependent on the accuracy of the requirement analysis which is the basis and source of software system development. However, manual preparation of software requirements specifications often leads to inconsistency with a business description, low efficiency, and communication difficulties with business personnel [19]. It is being observed that involving users extensively in the process of software requirements specification and design will lead to the creation of dependable and acceptable software systems [20].

Literature reveals that the concept of working in pairs has been observed to be very effective. This can be done in various phases of the software development life cycle (SDLC) phases such as programming in pairs, designing in pairs, testing in pairs, etc. Hence the work can be done effectively and efficiently. Pair programming is a collaborative programming approach in which two people work closely on the same task [21, 22]. While working together, the roles can be shifted among the collaborators. In this way, the work done will be efficient and effective [23]. A lot of research work has been done on pair programming [24-26]. In one of their research, Kude et. al [22] discussed the effectiveness of pair programming and stated that instead of solo programmers, the work done in pairs proved to be more effective. Furthermore, Chong et. al [23] have presented a comparison of paired and solo programmers. They have presented the analysis of 242 interruptions and have discussed team configurations as well. They have suggested pair programming to support better. Aguilar et. al presented an exploratory investigation that examines the impact of Belbin's Theory on the quality of software requirements development. A controlled experiment is employed to manipulate the factor of how development teams are integrated with students. The study findings demonstrate a statistically significant difference in product quality between teams composed of individuals with compatible roles and those that are randomly integrated [27]. In one of their research, Kim et. al [28] examined the productivity of pair programmers via a case study by doing repeat programming and reached various conclusions regarding novice-novice pairs against expert-expert pairs and solos as well. They state that novice–novice pairs against novice solos are much more productive than expert–expert pairs against expert solos. In addition to this, a comparison of Solo and Pair Programming has been conducted by Omer et. al. in terms of Flow Experience, Coding Quality, and Coding Achievement [29]. After conducting several solo and group experiments, it was concluded that pair programming is effective to be used in educational settings. An empirical study was conducted by Qamar et. Al [30] to investigate the usefulness of pair testing. The productivity of experimental groups was compared which revealed that the quality of the work was better in the case of pair testing. This is because while working, two people work at the same place to test the same feature at the same time. Moreover, they can exchange ideas and discuss with each other to improve the process further.

In another research conducted at the University of Castilla-La Mancha in Spain [31], the difference between solo and pair designing was studied. The results revealed that a good quality design document was produced as a result of pair designing. However, it has been observed that pair designing in various projects for experimentation took more time to complete the designing phase as compared to solo work. Swamidurai et al. [32] have examined the influence of Pairing on various phases of Software Development. They assessed whether Pairing is necessary for all phases by examining its impact on each phase of the software development life cycle. In a comprehensive study involving students as participants, they concluded that Pairing is equally advantageous during the design and testing phases as it is during the coding phase. In this work, we have evaluated the concept of pair documentation in which we conducted an experiment on one project and there was a single comparison conducted among a single person versus a single pair of participants for requirements specification. And the results were gathered to check the effectiveness of the pair documentation approach. Despite the importance of pairing and homogeneity between team members [28-35], the concept of pairing in requirements engineering activities like documentation has not been addressed very well. Moreover, the usefulness and efficacy of working in pairs while programming, testing, and designing, encouraged us to take the initiative of testing the practicality of pair documentation.

## 3 Research Methodology

There are different phases involved in our experiment. Fig. 1 shows our proposed research methodology and the following sections provide the details of each phase.

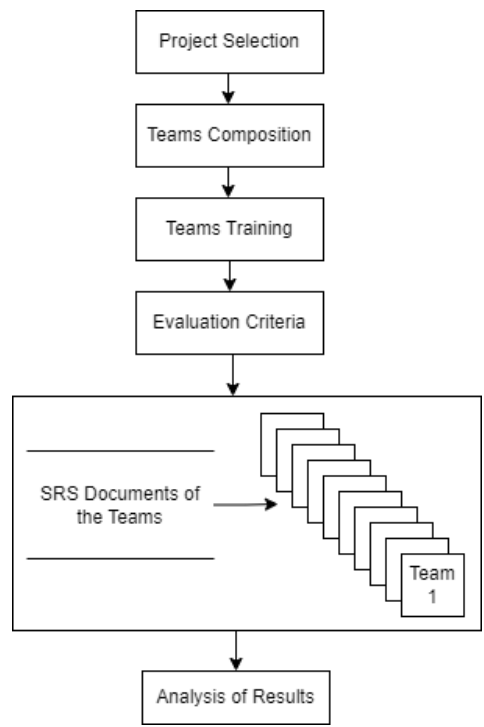

**Figure 1:** Research methodology

### 3.1 Project Selection

The first step in this research was an appropriate project selection. The project that we used for this experiment is a utility bill payment system (UBPS) that records all the transactions booked by call center agents. UBPS stores three different types of utility bill transactions i.e., booked, auto-debit, and mobile. Booked transactions need to be booked explicitly through a call center agent each time a utility bill needs to be paid. Auto-debit transactions, on the other hand, are booked once only through a call center agent. Once booked, the bill automatically gets paid every month as per the defined schedule. These transactions apply to utility bills but not to mobile prepaid bills. Mobile transactions are used for paying bills of mobile cellular companies. Once a bill transaction has been booked, it must be approved by the call center administrator. After a call center administrator approves a transaction, another call center administrator verifies it and annotates it with her remarks. Finally, the bill is marked by an online authentication (OLA) officer who pays the bill and sends an SMS to the customer who booked the transaction. The main reason for selecting this system was one author's in-depth familiarity with the domain of billing systems. This in-depth familiarity is needed later in the experiment when this author plays the role of a client providing requirements.

*3.2 Teams Composition*

In this study, 20 teams participated in the experiments. All the teams were asked to write down the requirements of the same system i.e. UBPS. One was the control group and the other was the experimental group. In the experimental group, 10 teams were working in pairs i.e. 20 participants. On the other hand, there was a control group in which 20 participants worked alone on the same project. Each team in the experimental group contained people of the same level i.e. qualifications, experience, and rank were the same. The control group also had 20 participants of the same level of expertise. A few teams consisted of students who are about to graduate or are in their last semester. Whereas some teams consist of participants who are on internships or doing the job. Hence are more experienced to work on requirements. Furthermore, the teams being made in pairs were also uniform in terms of knowledge and expertise and none of them is superior to the other.

*3.3 Training of Teams*

Before the teams started actual documentation, a training session was organized to deliver the requirements and guidelines of the experiment. According to the discussed guidelines, all requirements had to be written in the natural language (English) and at the same level with every minute detail in them. Every requirement will be documented in a manner that every requirement will have its unique requirement ID. All the functional requirements will have requirement ID as Req-001 onward and all the Non-Functional requirements will have requirement ID as Req-N-01 and so on. Requirements will be documented using MS Word (Fig 3 shows sample output as per the given guidelines). In addition to this, every member was responsible for his/her work, and they were not allowed to share any material related to the project with anyone. To get the requirements, one of the authors portrayed a client to generate a reference SRS which was not shared with any team but was only to do the comparison at the end of the experimental phase. The training was then provided to all the participants of the control and experimental groups, and everyone attended those sessions together to receive the same information together. These training sessions last over an hour consisting of 40 minutes to discuss the project and its requirements in detail and 20 minutes for the Q&A session. These training sessions were conducted not just to describe the project in brief, but they were productive in a manner that the proper training agenda was explained, functional and non-functional requirements were clarified, and the session also depicted how to write clear, complete, and concise requirements.

*3.4 Evaluation Criteria*

The next phase of this research was evaluation criteria to answer the following research questions:

1) Which group proved to be more productive?

2) Which group produced a better-quality SRS document?

The answer to the first question was found by measuring the productivity of the teams by evaluating the completeness of the SRS document in terms of functional and non-functional requirements. Functional requirements describe the functionality of a system in detail and mention how the system should behave in different situations. Non-functional requirements, on the other hand, deal with different quality or level-of-service aspects such as reliability, portability, usability, and performance. Completeness was judged by comparing requirements documented by the teams with the reference SRS document.

Productivity= Completeness of functional and Non-Functional Requirements/Effort in Person Hours  (1)

To answer the second question, the quality of SRS documents was judged by looking at the presence of ambiguity in the listed requirements and inconsistency in SRS documents. A requirement is considered ambiguous if it has multiple interpretations. Even though there are different types of ambiguities e.g. lexical, semantic, referential, etc. [36], we focused on lexical ambiguities only. We have studied different quality assessment options [37-39]. After that, we selected an automated ambiguity detector tool (proposed by Nigam et al [40]. ) that was used to detect the presence of lexical ambiguities in the SRS documents. This tool takes input from the requirements and configuration files. The requirements file contains a list of functional and non-functional requirements whereas the configuration file includes a list of some ambiguous words as shown in Table 1. This list is used as a reference point for ambiguity detection in SRS. We measured the inconsistency in requirements manually comparing every documented requirement with every other documented requirement in the same SRS. Fig. 2 shows the automated ambiguity detector tool used to detect the presence of any lexical ambiguities in the SRS. It takes as input the requirements file consisting of all functional and non-functional requirements and the configuration file consisting of a list of ambiguous words.

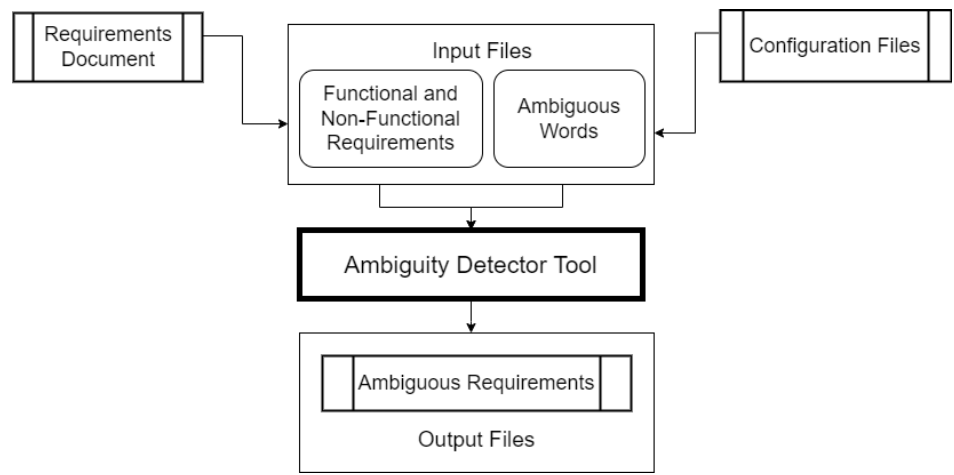

**Figure 2:** Ambiguity detector tool

Table 1 shows the list of ambiguous words (a sample is shown) which acts as a reference point for ambiguity detection. Moreover, while parsing requirements, it checks for the presence of all these words as mentioned in Table 1. After the parsing of the entire requirements, the results are generated as an output file. A sample output generated by this tool can be seen in Fig. 3.

**Table 1:** Sample List of ambiguous words

| Ambiguous Words | | |
|---|---|---|
| Accommodate | Adequate | And |
| As a minimum | Easy | Etc. |
| Robust | Can | May |
| Some | They | That |
| Always | None | Fast |
| Normal | Effective | This |
| Those | Never | Every |
| Capability to | Someone | Often |
| Capability of | Obviously | Support |
| Maximize | Therefore | Sufficient |

```
------------------------------------------------------------------------
This file contains the ambigious requirements of Team 1 (Experimental Group).
------------------------------------------------------------------------

Req_001   Every User should access the application using authorized username and password.
ambigious word : and

Req_002   User should enter the username and password on login page.

ambigious word : and
Req_003   If system authorized the username and password then user successfully access the application according to role.

ambigious word : be able to

Req-N-02   The messages should be encrypted for login communications, so others cannot get username and password from those messages.
ambigious word : can, and

Req-N-03   If a user wants to create an account and the desired user name is occupied, the user should be asked to choose a different user name.
ambigious word : and

Req-N-04   The system shall accommodate many users during the peak usage time
ambigious word : accommodate

Req-N-09   The results displayed in the list view should be user friendly and easy to understand.
ambigious word : easy, and

Total No of Requirements in SRS of Team 1 : 82

Total No of Functional Requirements in SRS of Team 1 : 68

Total No of Non-Functional Requirements in SRS of Team 1 : 14

Total No of Ambigious Requirements in SRS of Team 1 : 14

Total No of Ambigious Functional Requirements in SRS of Team 1 : 6

Total No of Ambigious Non-Functional Requirements in SRS of Team 1 : 8
```

**Figure 3:** Sample output

Table 2 shows the details of the experimental attributes. According to it, there were a total of 126 requirements out of which 100 are functional and 26 are non-functional requirements. This experiment involves a total of forty BS(CS) final-year students who were divided into twenty teams (10 teams of the experimental group and 10 teams of the control group). It took almost 2 months to complete this experiment from planning to results compilation and analysis, out of that 14 working days were given to students to write their SRS documents.

**Table 2:** Experimental details

| Attributes | Description |
|---|---|
| Domain of reference SRS document | Online Transactions |
| Total Requirements in Reference SRS document | 126 |
| Functional Requirements in Reference SRS document | 100 |
| Non-Functional Requirements in Reference SRS document | 26 |
| Participants Involved | Final Year Students of BS(CS) |
| No. of Participants Involved | 40 |
| No. of Teams | 20 |
| Time allowed to complete the task | 14 working days |
| Experimentation total time | 2 months |
| Training duration | 120 minutes (60 Minutes for each group) |
| Evaluation Criteria/Metrics used for document quality | % Ambiguity in requirements<br>% Inconsistency in requirements |
| Evaluation Criteria/Metrics used for document productivity | % Completeness in requirements |

### *3.5 Control and Experimental Group Documentation*

All the teams worked in parallel after this training session was over, for almost 14 working days to compile the SRS document for the system. The SRS document was taken from all the teams after the cut-off date.

### *3.6 Analysis of Results*

In this experiment, the final step was to compile and analyze the results. The productivity and quality of SRS documents were compared between the two groups. Table 3 and Table 4 show the summary of results concerning the completeness of the SRS document generated by both the experimental group and the control group respectively. It can be observed that almost all the teams of the experimental group were able to gather more functional requirements out of 100 functional requirements and 26 non-functional requirements as compared to the control group because they used a pair documentation procedure whereas the teams of the control group used traditional way to gather requirements. It can be evaluated that the experimental group proved to be more productive providing a complete SRS as both the teams were working on the same project and requirements were given in the same time frame. The whole experimentation generated significant results when quality requirements were gathered i.e. consistent and unambiguous requirements by the same teams. Therefore, in terms of ambiguity and inconsistency, it can be observed that the requirements identified by the control group were more ambiguous and inconsistent as compared to the requirements identified by the teams of the experimental group as shown in Tables 3 and 4. For instance, Team 1 of the experimental group gathered a total of 82 out of 126 total requirements, out of which 68 out of 100 were Functional Requirements (FR) and 14 out of 26 were Non-Functional Requirements (NFR). On the contrary, Team 1 of the control group was able to gather a total of 74 out of 126 requirements from which 64 were FR and 9 were NFR. This shows the experimental group to be more productive. Similarly, in terms of quality, it is evident that experimental group teams gathered more ambiguous and consistent requirements as compared to the control group teams. For example, 17 out of

100 Ambiguous Requirements (AR) and 93 out of 100 CR were gathered by experimental Team 1 as compared to 30 out of 100 AR and 82 out of 100 Consistent Requirements (CR) gathered by the control group teams for project 1.

**Table 3:** Comparison concerning completeness, ambiguity, and consistency for experimental group teams

| Sr# | Teams | Experimental Group Teams | | | | | | |
|---|---|---|---|---|---|---|---|---|
| | | Completeness | | | Ambiguity (%) | | Consistency (%) | |
| | | TR | FR | NFR | NAR (%) | AR (%) | CR (%) | ICR (%) |
| 1 | Team 1 | 82 | 68 | 14 | 83 | 17 | 93 | 7 |
| 2 | Team 2 | 96 | 81 | 15 | 75 | 25 | 81 | 19 |
| 3 | Team 3 | 55 | 43 | 12 | 68 | 32 | 73 | 27 |
| 4 | Team 4 | 105 | 89 | 16 | 89 | 11 | 67 | 33 |
| 5 | Team 5 | 68 | 57 | 11 | 63 | 37 | 58 | 42 |
| 6 | Team 6 | 89 | 72 | 17 | 92 | 8 | 87 | 13 |
| 7 | Team 7 | 98 | 84 | 14 | 58 | 42 | 61 | 39 |
| 8 | Team 8 | 101 | 91 | 10 | 87 | 13 | 83 | 17 |
| 9 | Team 9 | 97 | 84 | 13 | 81 | 19 | 76 | 24 |
| 10 | Team 10 | 69 | 57 | 12 | 76 | 24 | 68 | 32 |

TR=Total Requirements, FR=Functional Requirements, NFR=Non-Functional Requirements, AR=Ambiguous Requirements, Unambiguous Requirements, CR=Consistent Requirements, ICR=Inconsistent Requirements

**Table 4:** Comparison concerning completeness, ambiguity, and consistency for control group teams

| Sr# | Teams | Control Group Teams | | | | | | |
|---|---|---|---|---|---|---|---|---|
| | | Completeness | | | Ambiguity (%) | | Consistency (%) | |
| | | TR | FR | NFR | NAR (%) | AR (%) | CR (%) | ICR (%) |
| 1 | Team 1 | 73 | 64 | 9 | 70 | 30 | 82 | 18 |
| 2 | Team 2 | 85 | 71 | 14 | 68 | 32 | 72 | 28 |
| 3 | Team 3 | 93 | 79 | 14 | 57 | 43 | 85 | 15 |
| 4 | Team 4 | 62 | 57 | 5 | 89 | 11 | 62 | 38 |
| 5 | Team 5 | 95 | 83 | 12 | 78 | 22 | 74 | 26 |
| 6 | Team 6 | 49 | 42 | 7 | 66 | 34 | 65 | 35 |
| 7 | Team 7 | 89 | 76 | 13 | 91 | 9 | 75 | 25 |
| 8 | Team 8 | 102 | 87 | 15 | 87 | 13 | 79 | 21 |
| 9 | Team 9 | 63 | 54 | 9 | 53 | 47 | 68 | 32 |
| 10 | Team 10 | 88 | 73 | 15 | 69 | 31 | 59 | 41 |

TR=Total Requirements, FR=Functional Requirements, NFR=Non-Functional Requirements, AR=Ambiguous Requirements, Unambiguous Requirements, CR=Consistent Requirements, ICR=Inconsistent Requirements

Fig 4 and 5 show the comparisons in terms of Completeness of FR and NFR respectively, conducted by the experimental and control group simultaneously. It can be seen that requirements collected by the experimental group satisfy the completeness criteria for the majority of the projects, hence answering question 1 of our research.

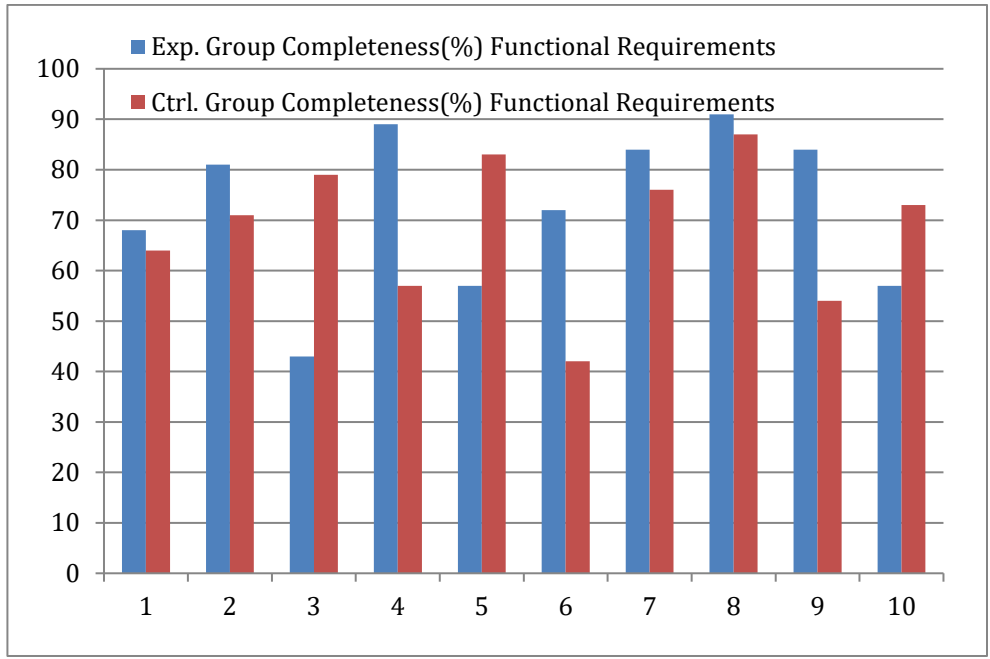

**Figure 4:** Comparison in terms of completeness of functional requirements

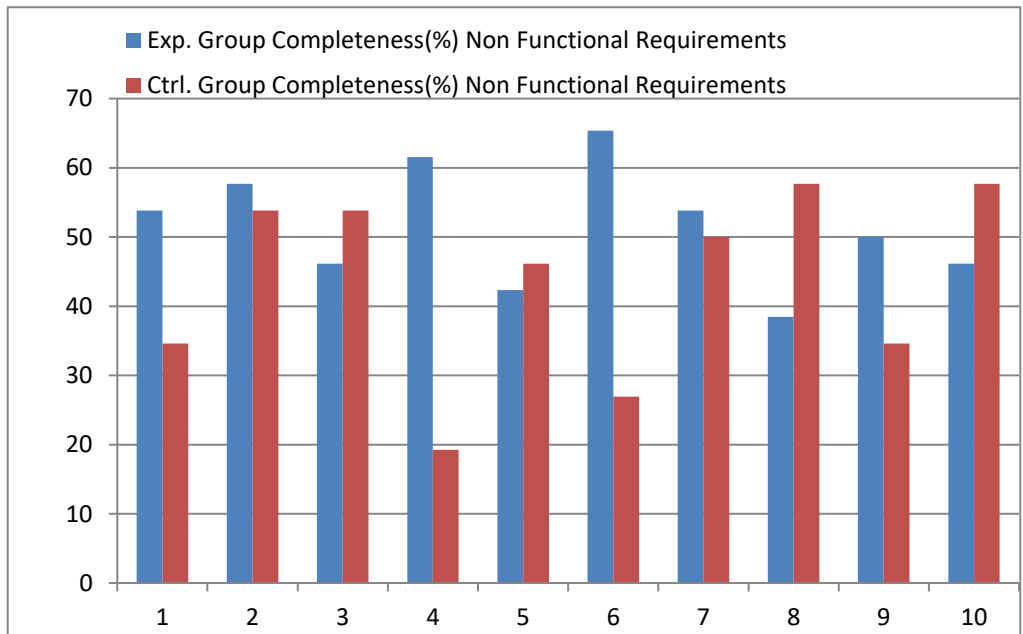

**Figure 5:** Comparison in terms of completeness of non-functional requirements

Figure 6 shows the percentage of ambiguous requirements collected by the teams of experimental and control groups. It can be seen that the percentage of ambiguous requirements gathered by the teams of the control group is more than the ambiguity value for Experimental group teams.

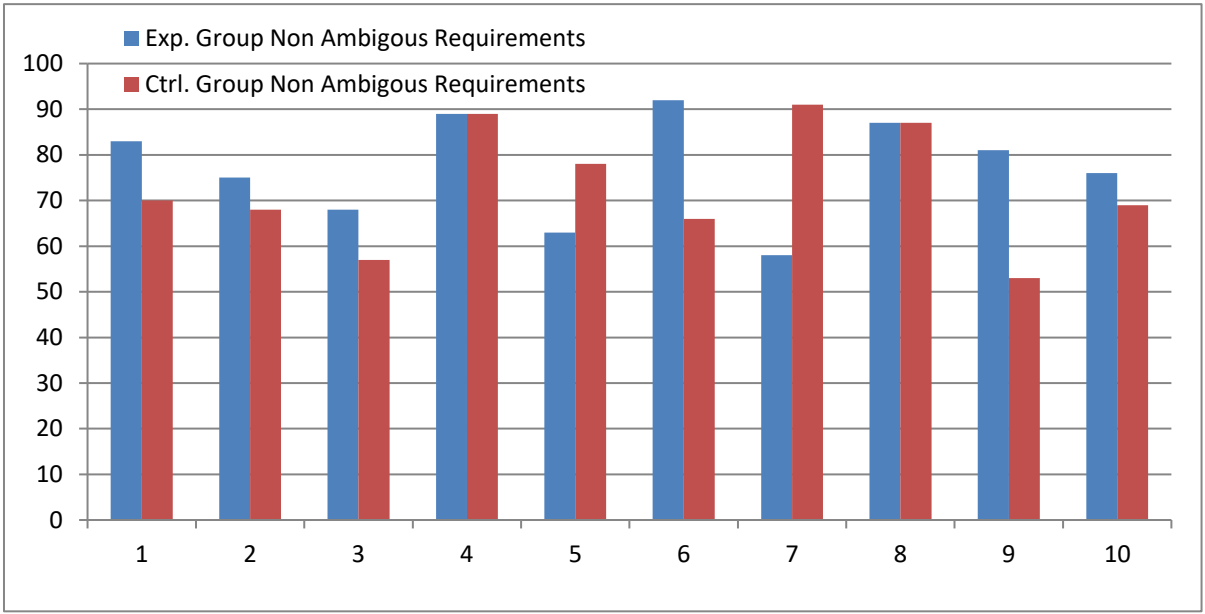

**Figure 6:** Comparison in terms of ambiguous requirements

Fig. 7 shows the comparison of the SRS documents in terms of the consistency of requirements. It is evident from the graphs that all the teams except Teams 3, 5, and 7 of the experimental group gathered consistent requirements. Therefore, these comparisons clearly show that the quality of requirements gathered by the experimental group teams using pair documentation is far better than those working alone. These results answer the second question of our research.

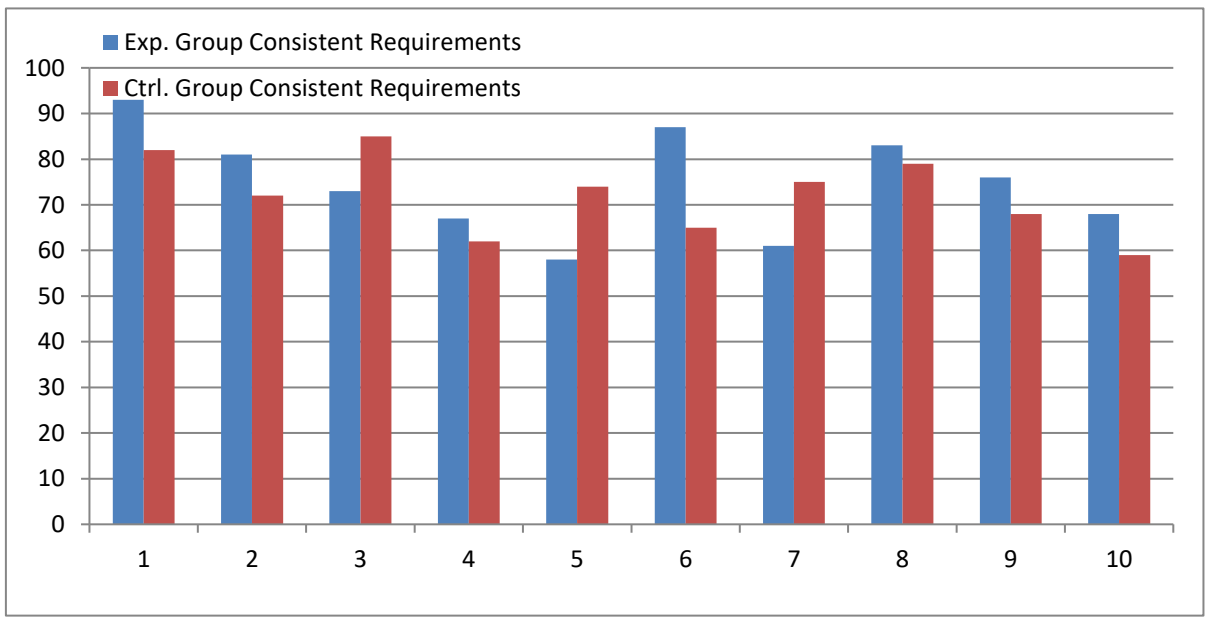

**Figure 7:** Comparison in terms of consistent requirements

As is evident from the above tables and figures, the teams from the experimental group (working in pairs) produce better quality SRS documents with less ambiguity and inconsistency in obtained requirements as compared to the control group. The same holds for team productivity. The rationale behind the experimental group's enhancement in productivity and SRS document quality is comprehensible since having two pairs of eyes is more effective than one in identifying issues and conducting real-time quality assurance of the work at hand.

**4 Threats to Validity**

Although our experimental findings favored pair documentation, it is imperative to recognize the potential impediments that may affect the interpretation of these results. Firstly, the divergent personality traits of individuals, such as agreeableness, intelligence, openness to experience, contentiousness, and inclination towards documentation and quality assurance, could have had an impact on the outcomes. For example, participants who possessed a higher level of agreeableness may have found the pair documentation approach more enjoyable and efficacious than those who did not share this personality trait. Secondly, it is pertinent to note that our subjects had no prior exposure to pair documentation and had only been familiar with traditional documentation practices. As a consequence, they may have been excessively enthusiastic about this innovative approach, which could have influenced the positive outcomes observed in our experiment. Nonetheless, it is plausible that with repeated exposure to pair documentation, the initial excitement may dwindle, leading to a reduction in the benefits observed. In conclusion, while our empirical results endorse the use of pair documentation, it is crucial to exercise prudence while interpreting these findings. Future research should endeavor to examine the impact of individual differences and the effects of repeated exposure to pair documentation to gain a more comprehensive understanding of its efficacy. The initial enthusiasm of participants towards pair documentation may have contributed to increased productivity, possibly due to the novelty of the approach, the presence of a collaborator, and perceived benefits. Pre-existing working relationships among pairs of student participants and professional experience among pairs may have also influenced their performance during the experiment. These factors should be considered in future research to better understand the effectiveness of pair documentation as a technique.

**5 Conclusion**

This research aimed to highlight the significance of pair documentation as compared to the traditional documentation approach. This is an empirical study conducted among two groups i.e. control group and the other one was experimental group, both consisting of ten teams. This is the only study that compared the impact of pair documentation involving multiple teams in two different groups. We were not able to find any other study similar to this research. The results of the experiments reveal that the teams using pair documentation (the experimental group) were more productive in terms of completeness of requirements i.e. they gathered more requirements in the given time. Moreover, the experimental group produced a better quality of requirements than the control group with less ambiguous and more consistent requirements. This research can be extended in several ways. Firstly, it can be replicated with different experimental settings e.g. different application domains, a greater number of subjects, a bigger project size, a higher experimental level, etc. This would also be worthwhile to explore the impact of pair documentation on other types of ambiguities (as we just explored the lexical ambiguity) e.g. semantic ambiguity, synthetic ambiguity, referential ambiguity, syntactic ambiguity, etc. Similarly, it would also be interesting to explore the impact of pair documentation on other activities of requirements engineering e.g. elicitation, analysis, validation, and documentation.

**Funding Statement:** The authors received no specific funding for this study.

**Conflicts of Interest:** The authors declare that they have no conflicts of interest to report regarding the present study.